# Probing the Intrinsic Optical Quality of CVD Grown MoS$_2$

Amina Zafar[1,†], Haiyan Nan[1,†], Zainab Zafar[2], Zhangting Wu[1], Jie Jiang[1], Yumeng You[2,*] and Zhenhua Ni[1,*]

[†]These authors contribute equally.

[1]*Department of Physics and Key Laboratory of MEMS of the Ministry of Education, Southeast University, Nanjing 211189, China.*

[2]*Ordered Matter Science Research Center, Southeast University, Nanjing 211189, China.*

*Corresponding authors: zhni@seu.edu.cn and youyumeng@seu.edu.cn

## ABSTRACT

The optical emission efficiency of two dimensional layered transition metal dichalcogenides (TMDs) is one of the most important parameter that affects their optoelectronic performance. Optimization of growth parameters of chemical vapor deposition (CVD) to achieve optoelectronics-grade quality TMDs is therefore highly desirable. Here, we present a systematic photoluminescence (PL) spectroscopic approach to assess the intrinsic optical and crystalline quality of CVD grown MoS$_2$. We suggest that the intensity ratio between PL measured in air and vacuum could be used as an effective way to monitor the intrinsic optical quality of CVD MoS$_2$. Low temperature PL measurements are also used to evaluate the structural defects in MoS$_2$ by defect-associated bound exciton emission, which is well correlated with the field effect carrier mobilities of MoS$_2$ grown at different temperatures. This work therefore provides a sensitive, noninvasive method to characterize the optical properties of TMDs, allow tuning of growth parameters for the development of optoelectronic devices.

**KEYWORDS** Transition metal dichalcogenides, MoS$_2$, chemical vapor deposition, photoluminescence, defect, mobility

# 1 Introduction

The rich and diverse physics with novel optical and electrical transport properties of transition metal dichalcogenides (TMDs) offer unprecedented opportunities for electronics and optoelectronics applications [1-4]. Monolayer TMDs, e.g. $MoS_2$, are direct-gap semiconductors exhibiting extraordinary light absorption and emission properties which make them promising candidates for optoelectronic devices such as photodetectors [5, 6], photovoltaics [7] and light emitters [8]. Chemical vapor deposition (CVD) is the most promising method to produce large-scale film or high quality single crystals of TMDs [9-11]. However, the as-grown CVD TMDs are highly prone to chalcogen deficiency due to high volatility of chalcogenides, and therefore contain an abundance of chalcogen vacancies. These defects activate non-radiative recombination channels, thereby quenching the intrinsic PL efficiency, and reduce carrier mobility by acting as scattering centers [12,13]. In the meantime, it has already been demonstrated that the PL efficiency of TMDs is strongly influenced by a lot of factors, e.g, doping, molecular adsorption, strain, and defects [14-19]. In particular, owing to the large variations in CVD grown TMDs such as grain size, crystal orientation and presence of defects, a fundamental understanding on the intrinsic PL efficiency based on material quality remains elusive.

Here, we present a systematic spectroscopic approach to evaluate the intrinsic optical quality of CVD grown $MoS_2$. The amount of structural defects in CVD grown single crystal $MoS_2$ of different crystal sizes and growth temperatures is probed by low–temperature defect-induced bound exciton emission, which is well correlated with the difference in PL intensities measured in air and vacuum. The structural defects in CVD $MoS_2$ would not only affect its optical quality, but also its electrical performance, as demonstrated by electrical transport measurements. Our results thus provide a systematic approach for monitoring the intrinsic qualities of TMD's materials and facilitating the development of optoelectronic devices.

## 2 Experimental

### 2.1 CVD Growth

Monolayer $MoS_2$ grains were grown by chemical vapor deposition on 300 nm $SiO_2$/Si substrates. The substrates were placed face down above ceramic boat containing $MoO_3$ (99.5%, Aladdin) and located in the heating zone of CVD furnace, whereas sulfur (99.5%, Alfa Aesar) was located upstream at a distance of 30 cm from $MoO_3$ and heated separately. The quartz tube was initially pumped down to a base pressure of $10^{-1}$ Pa and flushed with argon gas (99.995%) using 200 sccm repeatedly to remove oxygen content. The temperature was raised to 105°C and kept for 1 hr to remove water contamination. After that, the ramping rate was set at 15°C/min and approached to desired growth temperatures (600 – 900°C) with 10 sccm. The growth time was held to 5 min for all temperatures. After that, the system was cooled to specific temperature followed by rapid cooling by opening the furnace with 200 sccm argon gas. During the synthesis process the temperature of sulfur was kept at 230°C

### 2.2 Characterization and device fabrication

The Raman/PL measurements were carried out using a Horiba LabRAM HR800 micro-Raman system with 514.5 nm excitation, with an estimated laser spot size of 1 μm by using a 50x objective and the laser power is kept below 0.5 mW to avoid laser-induced heating. In order to record PL spectra in air and vacuum, the samples were placed in the same chamber to rule out the influence of glass window of chamber, and provided similar environmental conditions for all the measurements. Low temperature measurements were performed in a INSTEC HCP621V stage with mk1000 high precision temperature controller and LN2-SYS liquid nitrogen cooling system. The thickness of grown monolayer $MoS_2$ was confirmed by AFM (Asylum Research model: MFP-3D). For electrical measurements, the as-grown $MoS_2$ samples were transferred to fresh $SiO_2$/Si substrate by polymer assisted dry transfer method [20]. Successive steps included: the drop-casting of polypropylene carbonate (PPC) onto grown samples; baking for 30 min at $90^oC$; peeling off of the PPC/$MoS_2$ layer; and its transfer to the target substrate. Finally, the PPC was removed by acetone. The source and drain electrodes were patterned by standard electron beam lithography and followed by thermal evaporation of 5/50 nm of Ni/Au. The electrical transport measurements were measured by a Keithley 2612A SourceMeter under vacuum system (~1Pa) at room temperature.

## 3. Results and Discussion

By proper optimization and fine control of growth recipe, such as the amount and distance of feedstocks, growth temperature and duration, etc., highly crystalline $MoS_2$ has been synthesized over a broad temperature range by CVD method. Fig. 1(a) illustrates the growth dynamics of $MoS_2$ including the possible route of formation and diffusion of volatile suboxide ($MoO_{3-x}$) on the surface of substrate and the reaction of such suboxides with sulfur to initiate nucleation and formation of $MoS_2$ grains [21]. The optical and AFM images in Figs. 1(b) and 1(c) show that single crystals exhibit an equilateral triangular shape. The uniform color contrast suggests that the crystals are free from grain boundaries and adlayers [22]. The step height of each grain is ~0.8 nm which further confirms the formation of single layer $MoS_2$ [23]. Significant variations in grain size are observed for samples grown at different temperatures, and are presented in Figs. 1(d-g). At growth temperature of ~$600^oC$, the sublimation rate of precursors and generation of active species over the growth substrate is limited. As a result, the grown grains are preferentially small (average size of ~1μm) and have high nucleation density. As the growth temperature increases, active mobile species are produced and significantly diffuse on bare substrate, thereby causing the appearance of larger grains of sufficiently lower domain density. Notably, the largest grain growth (~20 μm) is observed at ~830$^o$C. On the other hand, the lateral dimensions of grains begin to shrink considerably when the growth temperature is too high (~900$^o$C), this may be attributed to the thermal etching of the sample and therefore impede the grain growth.

The Raman characteristics and intensity mapping of synthesized $MoS_2$ grains are presented in Electronic Supplementary Material Fig. S1, which reveals that the samples grown at different temperatures exhibit a good crystalline quality with spatial uniformity. However, the Raman results do not provide useful information in distinguishing the intrinsic optical property of $MoS_2$, due to the reason that Raman only probes the vibrational properties and the perturbation in crystal lattice, and is insensitive for detecting the relatively small variation in the properties of TMDs. On the other hand, PL process results from the radiative recombination of photoexcited electron-hole pairs, and is more sensitive to the structural defects, charged impurities and crystallinity of TMDs. Fig. 2(a) shows representative room-temperature PL spectra for single layer $MoS_2$ grown on $SiO_2$/Si substrate at different temperatures (600–900$^o$C). We also performed the integrated PL intensity mapping and observed uniform distribution of PL intensities across the whole $MoS_2$ grain (inset of Fig. 2(a)). The most interesting feature observed in Fig. 2(a) is the significantly increase of PL intensity as the growth temperature increases, with maximum value at 830$^o$C, and decreases with further increase in temperature. There is also a variation in line-width of grown $MoS_2$ at different temperatures (Electronic Supplementary Material Fig. S2). The narrowest line-width is found to be ~56 meV for 830$^o$C which is comparable to that of suspended samples of exfoliated MoS$_2$ [24]. These results are coincident with the change in the crystal gain size of MoS$_2$ with growth temperature. However, the quantum efficiency of TMDs in air is governed by certain constraints such as doping from various adsorbates (e.g. electronegative species O$_2$ and H$_2$O) and the existence of defects in crystal lattice. The former effectively transfers charges into/from MoS$_2$ and results in the transformation of different excitonic species, e.g. excitons into trions or vice versa, while the latter provides non-radiative recombination channels, sites for bound excitons, and adsorption sites for impurities/molecules.

The interaction of adsorbates with MoS$_2$ is physical adsorption and can be removed by vacuum pumping. A significant suppression in PL intensity as compared to its values measured in air is observed (Fig. 2(b)). To shed light on the interaction of adsorbates with MoS$_2$, Fig. 2(c) shows the PL spectra of MoS$_2$ grown at 830$^o$C obtained in air and vacuum (~0.1 Pa), respectively. The PL spectrum in vacuum can be fitted by two Lorentzian peaks instead of a single peak in air condition. The higher and lower energy emissions located at ~1.83 eV and ~1.80 eV, are termed as neutral (X$_0$) and charged exciton (X$^-$) respectively, while their difference is the binding energy of trion (E$_b$ ~30meV). These results are consistent with previous reports on MoS$_2$ [25,16] and other TMDs [26]. The remarkable difference in PL intensity between air and vacuum conditions can be interpreted as the change of excitonic states in MoS$_2$ due to change of doping concentration in different environments. In vacuum, the MoS$_2$ sample is heavily n-doped due to the presence of defects and unintentional doping of the substrate [25, 27], as clearly revealed in the transport curve shown in the inset of Fig. 2(c). Therefore, the PL of MoS$_2$ is highly dominated by the combination of excess electrons with X$_0$ to form stable X$^-$. While in air, the excess electrons get depleted with the adsorption of O$_2$ and/or H$_2$O molecules on MoS$_2$ surface, and thus switch the dominant PL process from trion recombination to exciton recombination. It is reported that the radiative recombination rate of exciton is much higher than that of trion [28], and this causes the strong difference in PL intensity of MoS$_2$ when measured in air and vacuum. Fig. 2(d) shows the PL intensities of single layer MoS$_2$ grown at different temperatures measured in air and vacuum. It is found that the highest intensities measured in air and vacuum are both from 830$^o$C sample, suggests that the difference in PL emission efficiency is related to the intrinsic quality of MoS$_2$ crystals. Fig. 2(d) also reveals another important feature that, in air, the PL intensity of the sample grown at 830$^o$C is ~27-fold higher than that grown at 600$^o$C, whereas in vacuum, it is only ~3.7-fold. Such phenomenon will be addressed later.

Next, we will focus our attention on defects in MoS$_2$. Basically, structural defects act as non-radiative recombination centers and can critically degrade the optical properties of CVD MoS$_2$. Therefore, the characterization of defect and defect engineering is important for electronic and optoelectronic applications of TMDs [29, 30]. For CVD-MoS$_2$, the non-uniform distribution of stoichiometry, such as varying degrees of sulfurization of MoO$_3$ on bare substrate, would lead to different amounts of defects within the crystal structure. These native defects are mainly comprised of point defects, grain boundaries and dislocations [13, 31]. Recently, study on the structural stability of different point defects in terms of their formation energies has been carried out using high-resolution Scanning Transmission Electron Microscopy (STEM) [12, 32]. Dark-field optical microscopy has also been employed to visualize the distribution of point defects in CVD grown MoS$_2$ on TiO$_2$/Ag substrate [33]. The S-vacancies are found to be the most abundant point defects in CVD MoS$_2$ and could introduce localized donor states with energies ~0.2 eV above the valence band maximum and ~0.3 eV below the conduction band minimum within the bandgap [19]. In such cases, the excitons in MoS$_2$ could bound to the defect site and activate defect-associated, bound-exciton emission. It has been reported that the intensity of such bound exciton emission can be used to estimate the numbers of defects in MoS$_2$ [34]. To further clarify the prominent optical quality of our grown MoS$_2$, low temperature PL measurements have been performed (Fig. 3(a)). Different intensities of defect-related bound-exciton emission (X$_b$) have been observed for samples grown at different temperatures. Fig. 3(b) shows the intensities of bound exciton (X$_b$) and free exciton (X$_0$) of MoS$_2$ plotted versus growth temperature. The lowest X$_b$ peak and the highest X$_0$ peak are observed at 830°C, which indicates the best condition for crystalline MoS$_2$ growth in our case. Traditionally, to increase crystalline quality and minimize intrinsic structural defects, high temperature CVD growth is favored for TMDs [23]. Nevertheless, above a certain temperature, growth conditions become unfavorable and may deteriorate material quality. With the identification of defects in MoS$_2$, it is now easy to understand the differences in PL intensities of MoS$_2$ grown at different temperatures. The PL quantum efficiency is given by:

$$\eta = k_r / (k_r + k_{nr}) \quad (1)$$

where $k_{nr} = k_{defect} + k_{relax} + k_{other}$

and $k_{nr}$ are the radiative and non-radiative recombination rates respectively, $k_{defect}$ represents the defect-induced non-radiative recombination, $k_{relax}$ corresponds to electron relaxation within the conduction and valence bands, and $k_{other}$ represents other non-radiative recombination processes, such as Auger recombination processes [35-36]. With the increase of defect concentration, e.g. from 830 to 600°C, $k_{defect}$ and ultimately $k_{nr}$ would greatly increase, causing the quenching of PL intensity.

Now we will focus on the different PL intensities of samples grown on different temperatures measured in air and vacuum, as shown in Fig. 2(d). This difference is also represented by the ratio between PL intensities measured in air and vacuum ($PL_{Air}/PL_{Vac}$), as shown in Fig. 4(a). It is found that the variation in $PL_{Air}/PL_{Vac}$ is closely related to the amount of intrinsic defects in grown $MoS_2$ at different temperatures, as indicated by the intensity ratio of $X_b$ and $X_0$ peaks ($I_{Xb}/I_{X0}$). We can explain this phenomenon by considering the recombination pathways of excitons and trions in the presence of defects (Fig. 4(b)). Defects are non-radiative recombination centers for both excitons and trions, labeled by red and yellow arrows respectively in Fig. 4(b). Therefore, according to Eq.(1), the intensities of both $X_0$ and $X^-$ should decrease with the increase of the amount of defects. However, the radiative recombination rate of excitons ($k_r$) is much higher compared to that of trions ($k_{r-}$) [28]. Consequently, a decrease in exciton intensity is much more significant as compared to that of trions with the presence of defects. Since PL spectrum of $MoS_2$ in air is dominated by excitons, while that of $MoS_2$ in vacuum is composed of both excitons and trions (with trions as the dominant component), the decrease of PL intensity in air is more pronounced as compared to that in vacuum with the increase of structural defects. This will finally result in a larger $PL_{Air}/PL_{Vac}$ at 830°C as compared to other temperatures, since $MoS_2$ grown at 830°C contains the least amount of defects. It should be noted that the defect states in Fig. 4(b) are not only non-radiative recombination centers, but they can also bind with excitons and result in the appearance of bound exciton emission, as shown by the $X_B$ peak observed in Fig. 3(a). In a short summary, our experimental approaches to measure low temperature bound exciton emission, and also room temperature PL intensity under air and vacuum, i.e. the ratio ($PL_{Air}/PL_{Vac}$), provide effective ways to assess the intrinsic optical quality of $MoS_2$ and other TMDs materials. It is worth noting that the ratio ($PL_{Air}/PL_{Vac}$) is not affected by factors such as experimental conditions and substrates, as compared to the direct use of PL intensity in ambient air, which make it useful to assess the optical quality of TMDs.

Previous studies have shown that the electrical performance of $MoS_2$ is highly affected by charged impurities [37, 27], short-range defects [38, 39], S-vacancies/grain boundaries [12, 22], trapped charges at the interface of $MoS_2$ and oxide dielectrics [40] and variation in stoichiometry within the crystal [41], resulting in considerable fluctuation in mobility from device to device. To evaluate the electrical quality of as-grown single layer $MoS_2$, samples were transferred onto fresh 300 nm $SiO_2$/Si substrates by polymer assisted dry transfer method (see section 2.2). The back-gated FET devices were fabricated on samples grown at five different temperatures, i.e. 650, 700, 750, 830 and 900°C. The source and drain electrodes were patterned by standard electron beam lithography followed by thermal evaporation of 5/50 nm of Ni/Au. In order to explore the consistency in electrical performance, 5-10 devices have been tested for samples grown at each growth temperature. Fig. 5(a) shows the output characteristics ($I_{DS}$-$V_{DS}$) at varying $V_g$ for a typical $MoS_2$ grain grown at 830°C. The nearly linear and symmetric curves confirm that Ohmic contacts are achieved at the source-drain contacts and there is an increase in drain current with increasing positive gate voltage, which is n-type semiconductor behavior consistent with reported literature [21-23]. The transfer characteristics ($I_{DS}$ -$V_g$) of the same device at different $V_{DS}$ are shown in Fig. 5(b). The on/off current ratio lies within the range of ~$10^5$ for gate voltages (-60 to 60 V). To establish the comparison, transfer curves were taken for samples grown at different temperatures (Fig. 5(c)). The field effect mobility was estimated using the relation:

$$\mu_{FET} = (dI_{DS}/dV_g)(L/WCV_{DS}) \quad (2)$$

where $dI_{DS}/dV_g$ is the slope, L and W are the channel length and width respectively. The capacitance per unit area between the channel and the back gate is estimated to be ~$1.15 \times 10^{-8}$ Fcm$^{-2}$ ($C_i = \varepsilon_0\varepsilon_r/d_{ox}$, where $\varepsilon_0$ is permittivity of free space, $\varepsilon_r$~ 3.9 and $d_{ox}$= 300 nm). The distribution of mobilities of the measured devices is plotted in the inset of Fig. 5(c). The averaged values of 1.5 ± 0.2, 16.0 ± 4.2, 17.4 ± 2.7, 19.9 ± 5.4 and 13.2 ± 3.8 cm$^2$/Vs were observed for samples grown at 650, 700, 750, 830 and 900°C, respectively. The highest mobility of the sample grown at 830°C is consistence with the optical characterization results that it contains the least amount of structural defects. Moreover, the correlation between FET mobility and $X_b/X_0$ is established in Fig.5(d). It should be noted that samples grown at different temperatures present obvious offset in the mobility, which is due to the different amount of charged impurities adsorbed on MoS$_2$ surface during growth as well as transfer process. Defects e.g. vacancies are active centers for molecular adsorption and chemical functionalization, as a consequence, the interaction of adsorbed charged impurities on defective MoS$_2$ would be much stronger than those on perfect MoS$_2$ lattice. Nevertheless, within the same substrate and samples grown at same temperature, there is a good correlation between the mobility and the intensity ratio of $X_b$ and $X_0$ peaks. This strongly suggests that the amount of intrinsic defects in CVD grown MoS$_2$ as probed by optical measurements do play an important role in affecting its electrical performance, i.e. scattering carriers and reduce its mobility.

## 4. Conclusions

We have adopted low temperature PL spectroscopy to investigate the amount of structural defects in CVD grown MoS$_2$, and found that it is well correlated with the PL efficiency measured in air and vacuum. We also suggest that the intensity ratio between PL measured in air and vacuum could be used as an effective way to monitor the intrinsic optical quality of CVD MoS$_2$. A correlation between electrical and optical properties in the form of mobility and defect associated PL intensity is also established. Our results provide a systematical approach to monitor the intrinsic optical and electrical properties of TMDs and could help on the future modulation of the properties of the materials.

**Acknowledgements**

This work was supported by NSFC (61422503, 21541013, and 61376104), Natural Science Foundation of Jiangsu Province (BK20150596), the open research funds of Key Laboratory of MEMS of Ministry of Education (SEU, China), and the Fundamental Research Funds for the Central Universities.

**Figures**

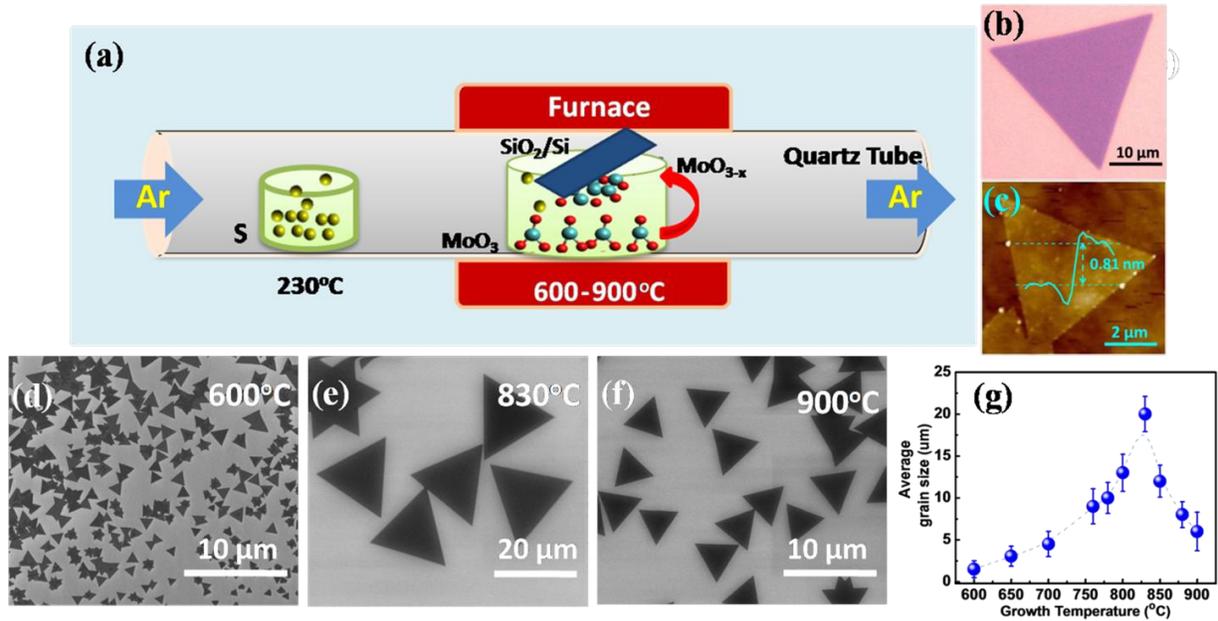

**Figure 1.** (a) Experimental setup for CVD growth of single layer $MoS_2$ on $SiO_2$/Si. $MoO_3$ is placed inside the heating zone of furnace and sulfur is loaded upstream outside the furnace and heated separately. (b) Optical and (c) AFM images of $MoS_2$ grain. (d-f) Variation in grain size of $MoS_2$ grown at different temperatures, as measured by scanning electron microscope. (g) Statistical data of grain size at different growth temperatures.

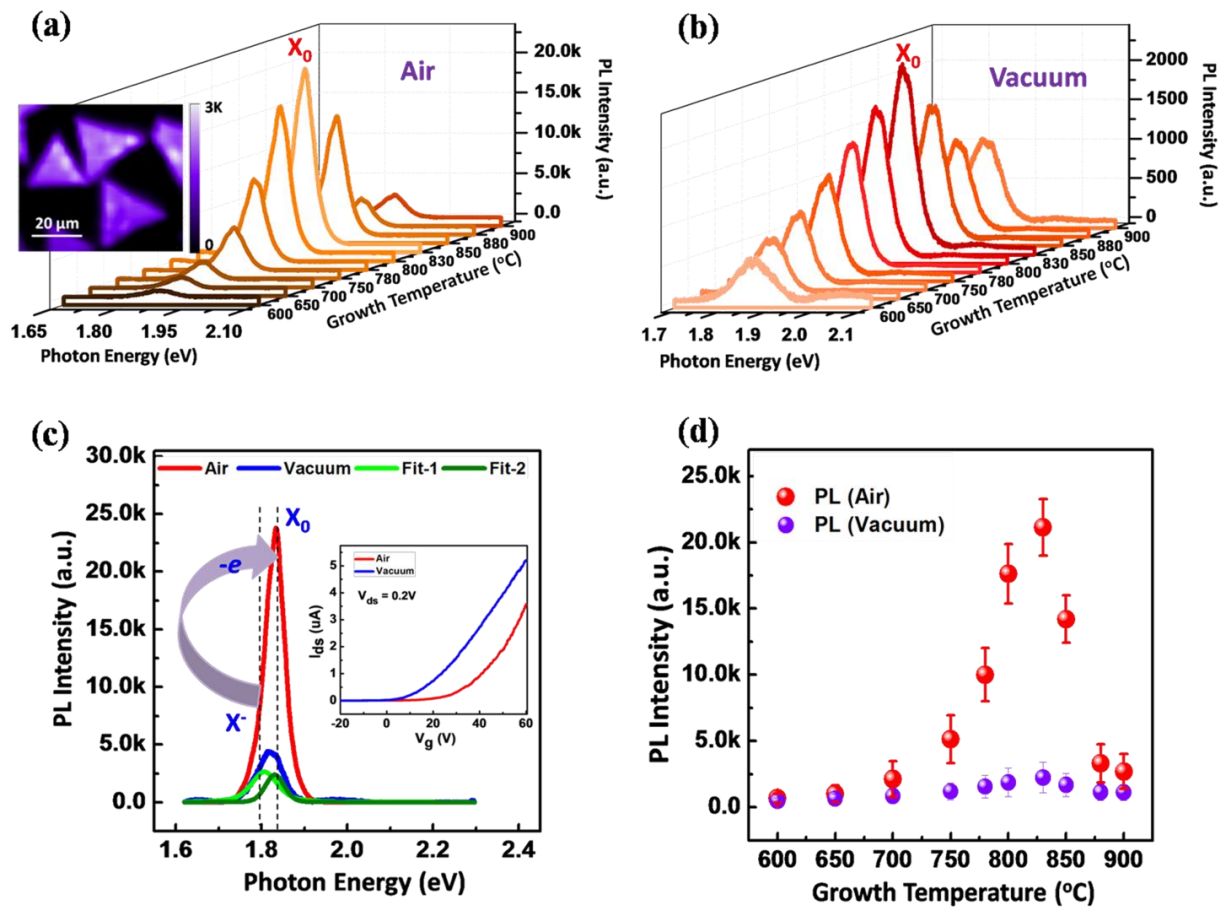

**Figure 2.** (a-b) PL spectra of single layer $MoS_2$ grown at different temperatures in air and vacuum. Inset is the PL image of single layer $MoS_2$ grains grown at 830°C over an area of 20 x 20 μm². (c) The PL spectra of a single layer $MoS_2$ grain grown at 830°C measured in air and vacuum. Inset is the transfer curves of single layer $MoS_2$ grown at 830°C measured in air and vacuum. (d) Comparison of average PL intensities of $MoS_2$ grown at different temperatures measured in air and vacuum.

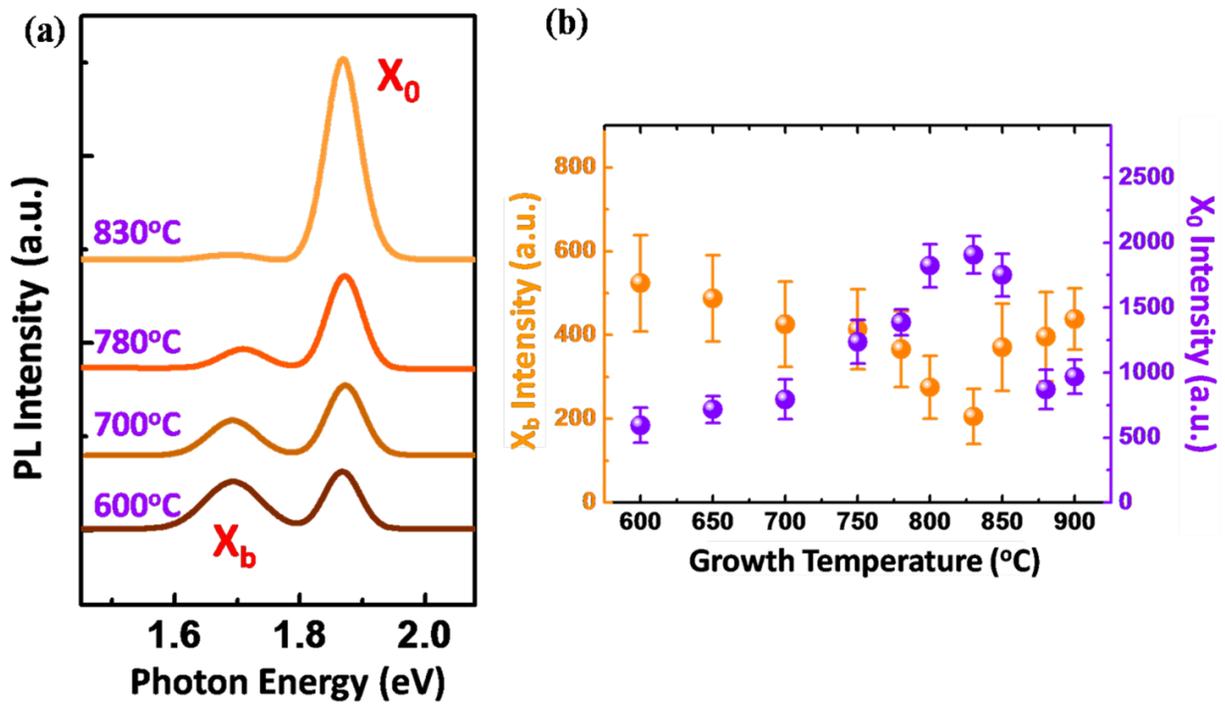

**Figure 3.** (a) PL spectra of single layer MoS$_2$ grown at 600, 700, 780 and 830°C measured at 83K. (b) Comparison of intensities of bound exciton peak (X$_b$) and exciton intensity peak (X$_0$) measured at 83K, as a function of growth temperatures.

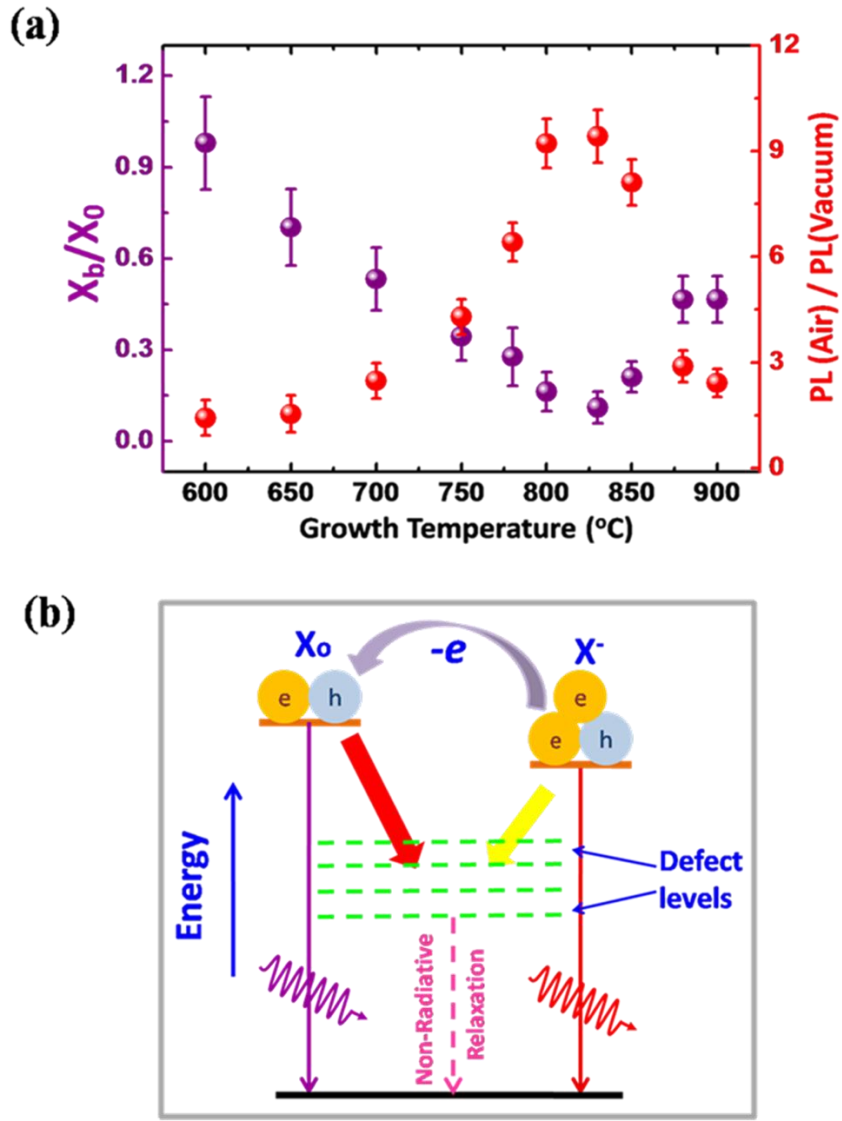

**Figure 4**. (a) Correlation of ratios ($X_b/X_0$) and PL(air)/PL(vac) of single layer $MoS_2$ as a function of growth temperatures. (b) The recombination pathways of excitons ($X_0$) and trions ($X^-$) in the presence of defects.

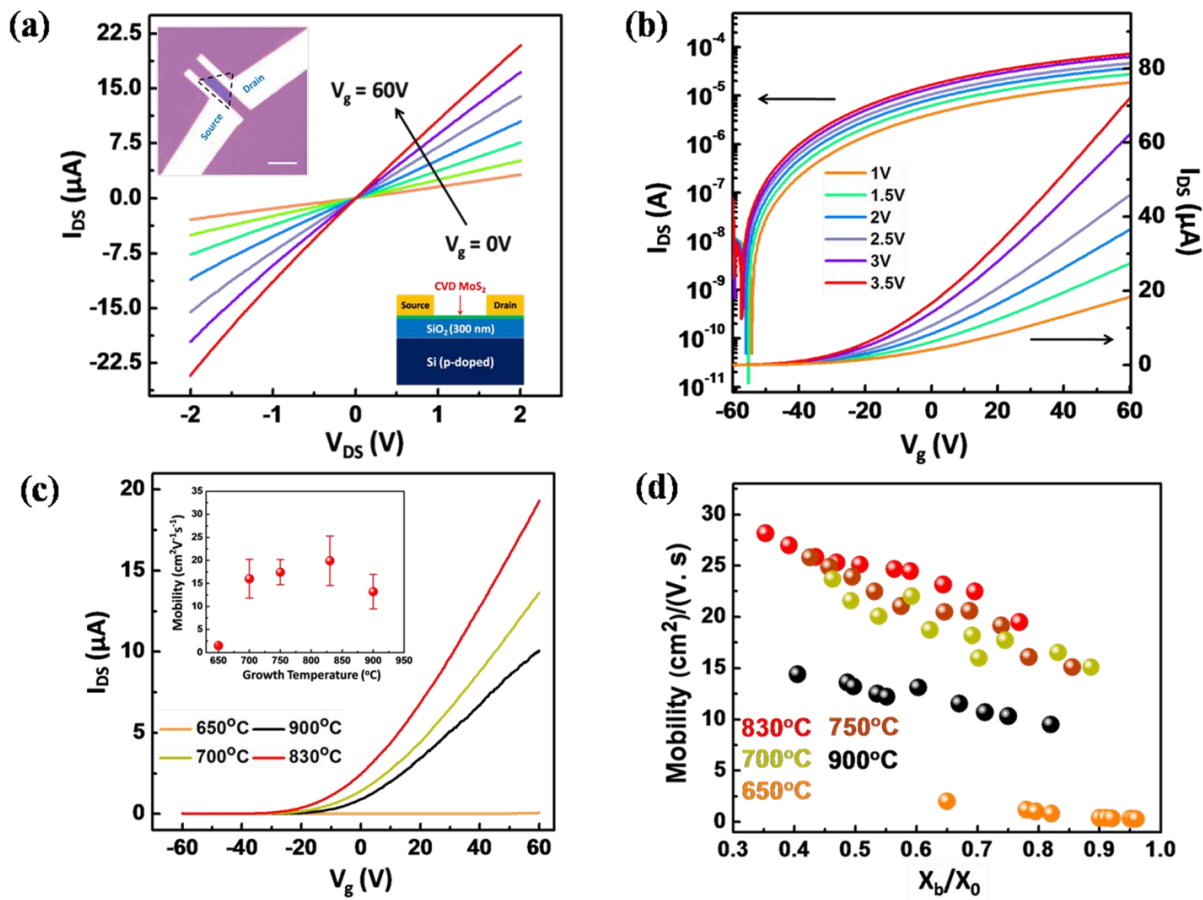

**Figure 5.** (a) Output Characteristics of $MoS_2$ FET grown at 830°C (back gate voltage sweeping from 0-60V in 10V steps). Inset is the optical image of the $MoS_2$ device. (b) Transfer characteristics of same FET (source-drain voltage changing from 1-3.5 V in 0.5V steps). (c) Transfer Characteristics of $MoS_2$ grown at different temperatures. Inset is the average carrier mobility of samples grown at different temperatures. (d) Correlation of carrier mobility and $X_b/X_0$ at different growth temperatures.

**Electronic Supplementary Material**

**Raman spectra of MoS$_2$ grown at different temperatures**

The Raman spectra of single layer MoS$_2$ grown at different temperatures are shown in Figure S1. The E$_{2g}$ mode arises from the in-plane vibration of an S–Mo–S layer whereas A$_{1g}$ comes from out-of-plane vibration of S-atoms [S1]. The uniformity of grown monolayer grains are further confirmed by Raman image integrated from the intensity of A$_{1g}$ mode. By precisely fitted the respective spectra at each growth temperature, the position of E$_{2g}$ and A$_{1g}$ are located within a range of ~384 - 385.5 cm$^{-1}$ and ~404 - 406 cm$^{-1}$ respectively, and the frequency difference of the two peaks are all ~20 cm$^{-1}$, in good agreement with the exfoliated and CVD monolayer MoS$_2$ [S1, S2]. Furthermore, the full width at half-maximum (FWHM) of E$_{2g}$ at each growth temperature is also extracted and found in the range of ~2.2 - 3.5 cm$^{-1}$, suggesting the high quality of CVD grown monolayer MoS$_2$ [S2].

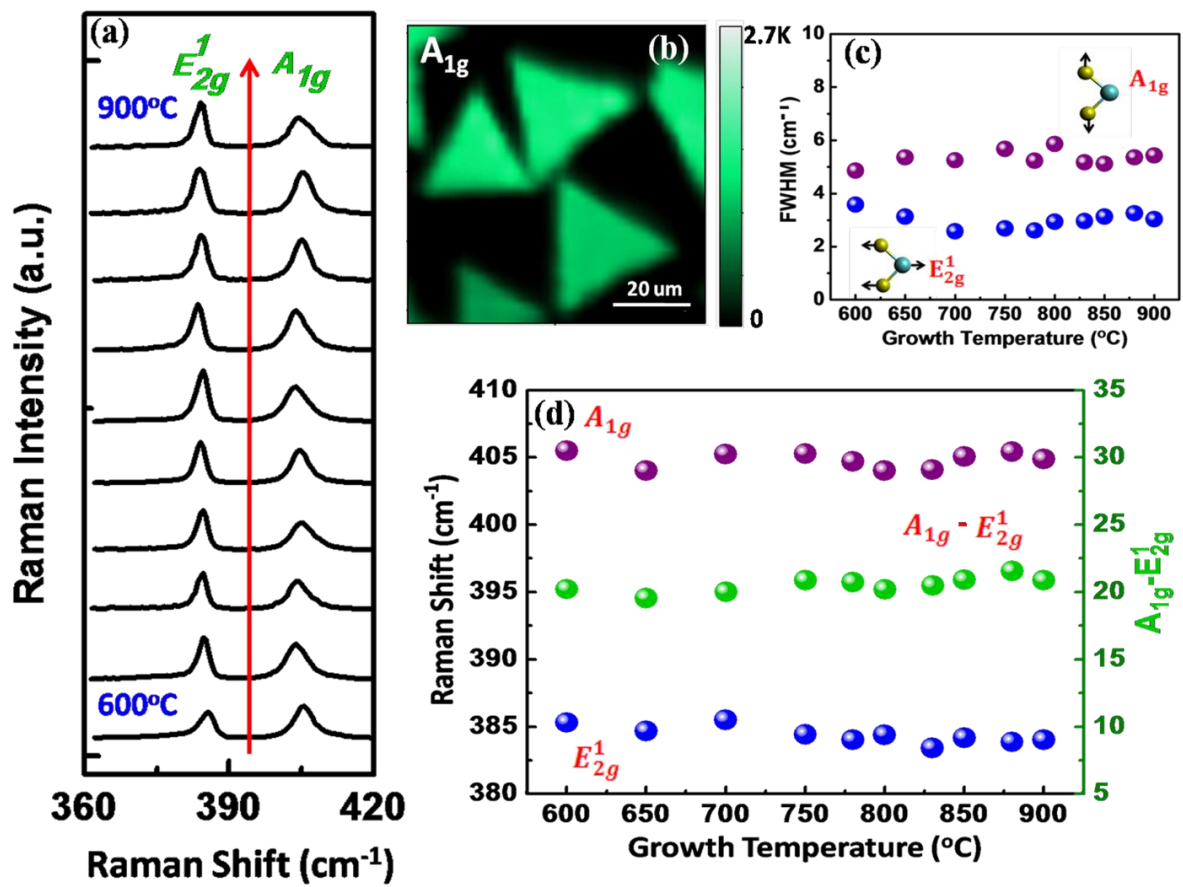

**Figure S1.** (a) Raman Spectra of single layer $MoS_2$ grown at different temperatures (b) Raman intensity image of $A_{1g}$ mode over an area of 20 x 20 $\mu m^2$ (c) The full width at half maximum (FWHM) of Raman modes at different temperatures. (d) The frequencies of $A_{1g}$ and $E_{2g}$ modes, as well as their frequency difference as a function of growth temperature.

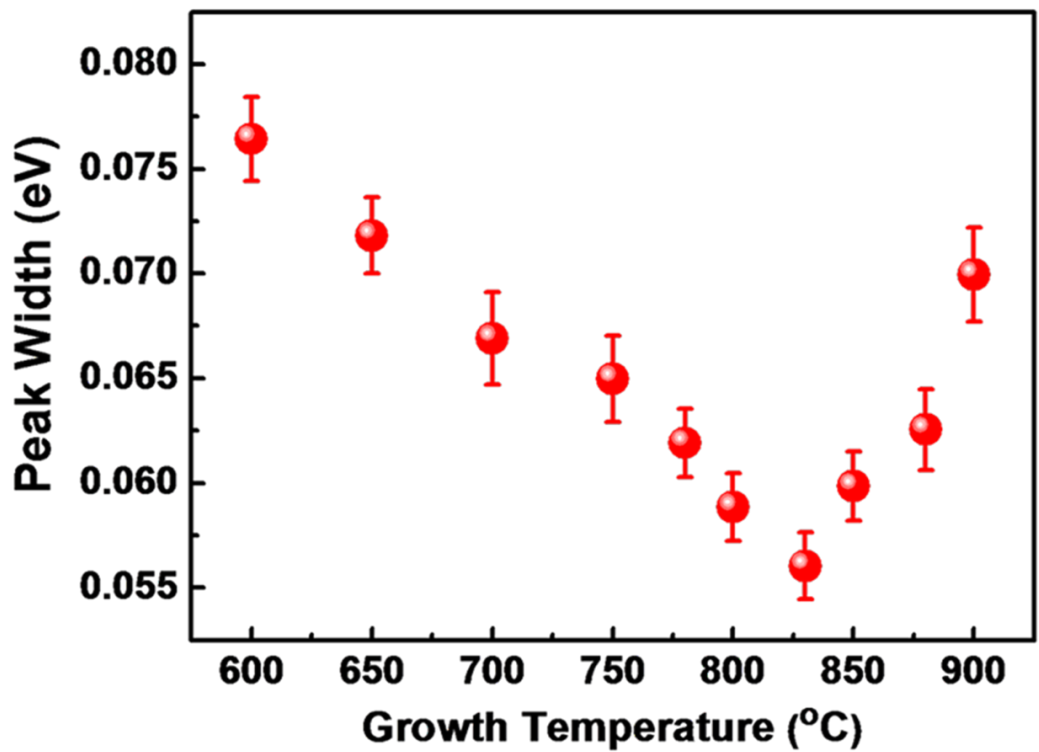

**Figure S2**. FWHM of PL peaks of single layer MoS$_2$ grown at different temperatures.